\documentclass[preprint,numafflabel, sort&compress]{elsarticle}
\usepackage[top=2cm, bottom=2cm, outer=2.5cm, inner=2.5cm]{geometry}
\usepackage[english]{babel}
\usepackage{lineno,hyperref}
\usepackage{multicol}

\usepackage{mathtools}
\usepackage{amsmath}
\usepackage{amssymb}
\usepackage{bbm} 
\usepackage{widetext} 
\usepackage{amsthm}
\usepackage{multirow}

\usepackage{longtable}
\usepackage{graphicx}
\usepackage{caption}
\usepackage{array}
\usepackage{float}
\usepackage{ragged2e}
\usepackage{graphbox} 
\usepackage{booktabs} 
\usepackage{colortbl} 
\usepackage{soul}
\usepackage[usernames,dvipsnames,svgnames,table]{xcolor}
\usepackage{hyperref}

\DeclareMathOperator{\Imag}{Im}

\journal{...}
\modulolinenumbers[5]

\begin{document}
	

	\begin{frontmatter}
		
		

		\title{Boundary conditions and infrared divergences}
		\author{Lissa de Souza Campos\footnote{lissa.desouzacampos@unipv.it}$^{\dagger\star}$}
		\author{Claudio Dappiaggi\footnote{claudio.dappiaggi@unipv.it}$^{\dagger\star}$}
		\author{Luca Sinibaldi\footnote{luca.sinibaldi01@universitadipavia.it}$^{\dagger\star}$}
		\address{$^{\dagger}$Dipartimento di Fisica, Universit\`a degli Studi di Pavia, Via Bassi, 6, 27100 Pavia, Italy}
		\address{$^{\star}$Istituto Nazionale di Fisica Nucleare -- Sezione di Pavia, Via Bassi 6, 27100 Pavia, Italy}

		\begin{abstract}
			We review the procedure to construct quasi-free ground states, for real scalar fields whose dynamics is dictated by the Klein-Gordon equation, on standard static Lorentzian manifolds with a time-like boundary. We observe that, depending on the assigned boundary condition of Robin type, this procedure does not always lead to the existence of a suitable bi-distribution $w_2\in \mathcal{D}'(M\times M)$ due to the presence of infrared divergences. As a concrete example we consider a Bertotti-Robinson spacetime in two different coordinate patches. In one case we show that infrared divergences do not occur only for Dirichlet boundary conditions as one might expect a priori, while, in the other case, we prove that they occur only when Neumann boundary conditions are imposed at the time-like boundary.
		\end{abstract}
		
		
		
	\end{frontmatter}
	

	
	\section{Introduction}\label{Sec: Introduction}
	
	In the past decade, we have witnessed a steadily growing interest towards the interplay between boundary conditions and the construction of free and interacting quantum field theories living on a large class of backgrounds possessing a (conformal) timelike boundary. From a physical viewpoint the reasons are several ranging from the desire of a better understanding of the structural aspects of a quantum field theory on asymptotically AdS spacetimes to the necessity of a thorough investigation of newly discovered phenomena, among which remarkable are the anti-Unruh and the anti-Hawking effect, see {\it e.g.} \cite{Brenna:2015fga, Henderson:2019uqo, DeSouzaCampos:2020ddx, DeSouzaCampos2022wsp} .
	
	In this endeavor, even when focusing on the simple scenario of a free, scalar field theory, one realizes from the very beginning that the presence of a timelike boundary entails already at a classical level a sharp difference from the standard formulation when the underlying background, say $(M,g)$, is globally hyperbolic. As a matter of fact, if the dynamics is ruled by a Klein-Gordon operator $P=\Box_g-m^2-\xi R$, see Equation \eqref{Eq: KG}, the associated initial value problem has to be supplemented with the assignment of a suitable boundary condition at $\partial M$. For several years, it has been customary to focus the attention only on that of Dirichlet type, since it has the distinguished property of being always admissible, in the sense that it entails not only that the underlying Cauchy problem is well-posed, but also that $P$ admits unique advanced and retarded fundamental solutions, which are in turn the building blocks to implement the canonical commutation relations at a quantum level. 
	
	Yet, in the past few years it has been highlighted that under suitable constraints on the parameters of the theory, in particular $m$ and $\xi$, one has the freedom of assigning to a scalar field theory a much larger class of admissible boundary conditions, among which noteworthy are those of Robin type, see {\it e.g.} \cite{Dappiaggi:2016fwc, Dappiaggi:2018jsq, deOliveira:2022gwd, deOliveira:2023qxe, Namasivayam:2022bky, Morley:2020zcd, Morley:2020ayr, Garbarz:2017wzv, deSouzaCampos:2021awm}. This realization has prompted the beginning of several research projects aimed at scrutinizing both the structural aspects and the physical properties of the underlying quantum counterpart. In this respect one of the main questions which needs to be addressed concerns the identification of a class of physically admissible quantum states. When the underlying spacetime is globally hyperbolic, it is widely accepted that one needs to restrict the attention to the so-called {\em Hadamard states} \cite{Khavkine:2014mta}. Once more denoting by $(M,g)$ the underlying background this entails that one needs to identify a bi-distribution $\omega_2\in\mathcal{D}^\prime(M\times M)$, which is a solution to the underlying equation of motion and whose antisymmetric part coincides up to a multiplicative constant to the difference between the advanced and the retarded fundamental solutions and it abides by a suitable constraint on its singular structure.
	
	At a physical level this establishes a condition on the ultraviolet behaviour of the quantum state which has far reaching consequences, among which it is worth recalling the existence of a covariant notion of Wick polynomials and the finiteness of the quantum fluctuations of all observables, see {\it e.g.} \cite{Fewster:2013lqa}.
	
	In view of these remarks, in the past few years, much effort has been devoted to finding a suitable generalization of the Hadamard condition whenever the underlying background $(M,g)$ admits a timelike boundary. Thanks to the effort of several research groups, it is nowadays understood what it is a good notion of Hadamard states in this framework and several explicit examples has been constructed, see {\it e.g.} \cite{Dappiaggi:2016fwc, Dappiaggi:2018xvw, Bussola:2017wki, deSouzaCampos:2020bnj}. Yet, in all these works, the focus has always been on the one hand on the control of the ultraviolet behaviour of the underlying two-point correlation function. On the other hand, barring some very specific examples, it has always been assumed that $(M,g)$ is stationary, since the existence of a timelike Killing field, entails that one can focus the attention either on ground or on thermal/KMS states as natural candidates for being of Hadamard form.
	
	In almost all scenarios considered, a great deal of attention has always been given to controlling the underlying ultraviolet behaviour. Yet almost no attention has been given to the possibility that infrared divergences can occur and that they are connected to the choice of an underlying boundary condition. The main goal of this paper is to warn about this possibility which, from a technical viewpoint, is tantamount to the existence of an obstruction in finding in the first place a suitable bi-distribution $\omega_2\in\mathcal{D}^\prime(M\times M)$ out of which one can define a desired, quantum state.
	
	More precisely, we shall proceed as follows. In the next section we shall recall succinctly and in great generality how one can construct on a globally hyperbolic spacetime with a timelike boundary $(M,g)$ a two-point correlation function associated to a ground state for a real, massive scalar field. This part of the paper is based on existing literature and its main goal is mainly to highlight were a potential infrared divergence can occur in the most general scenario, so to emphasize that this obstruction is always a potential threat.  Subsequently in Section \ref{Sec: Bertotti-Robinson} we focus the attention on a specific scenario, namely a real, massive scalar field on the four-dimensional Bertotti-Robinson spacetime \cite{Bertotti, Robinson}, here presented in two different coordinate patches. On the one hand this background is relevant at a physical level since it is both an exact solution of the Einstein-Maxwell's equations and it represents a good approximation of the near horizon geometry of Reissner-Nordstr\o m black hole, see {\it e.g.} \cite{Conroy2021aow}. On the other hand, the background isometries are such to allow us to give a detailed, analytic construction both of the underlying fundamental solutions with Robin boundary conditions. Subsequently we shall turn our attention to the construction of the two-point correlation functions of an underlying ground state and we shall show explicitly that, in one of the coordinate patches considered, only with Dirichlet boundary conditions, no infrared singularity is present. On the contrary, in the second coordinate patch, this pathology occurs only if we consider Neumann of boundary conditions, hence allowing in this case also those of Robin type. 	
	
	\section{General Construction}\label{Sec: General Construction}
	In this section we review succinctly the construction of the two-point correlation function associated to the ground state of a real, massive scalar field, so to highlight at which stage a potential infrared singularity might occur and the relation between its insurgence and the underlying boundary condition. Henceforth with $(M,g)$ we refer to a globally hyperbolic, oriented and time-oriented Lorentzian manifold with timelike boundary of dimension $\dim M=n\geq 2$, as formalized in \cite{Ake-Flores-Sanchez-18}.
	
	In addition, we assume that $(M,g)$ is standard static, namely it is isometric to $\mathbb{R}\times\Sigma$ with line element 
	$$ds^2=-\beta dt^2 + h,$$
	where $t\in\mathbb{R}$ plays the r\^{o}le of time coordinate along the $\mathbb{R}$-direction, while $\beta$ is a positive, time-independent, smooth function. In addition $h$ is a time-independent Riemannian metric on $\Sigma$. As a consequence the boundary $\partial M\simeq\mathbb{R}\times\partial\Sigma$ is also naturally endowed with a static metric. On top of this class of backgrounds we consider a real scalar field $\Phi:M\to\mathbb{R}$ whose dynamics is ruled by 
	\begin{equation}\label{Eq: KG}
		P\Phi=\left(\Box_g-V\right)\Phi=0,
	\end{equation}
	where $\Box_g$ is the d'Alembert wave operator built out of the underlying metric $g$, while $V$ is a time-independent potential, which we assume to be smooth in the interior of $M$. A rather common choice for such potential is $V=m^2+\xi R$ where $m\geq 0$ plays the r\^{o}le of mass parameter while $\xi\in\mathbb{R}$ denotes an arbitrary coupling to the scalar curvature $R$. Since $M$ possesses a timelike boundary, the solutions to Equation \eqref{Eq: KG} cannot be constructed only by assigning initial data on a Cauchy surface $\Sigma_{t_0}\simeq\{t_0\}\times\Sigma$, $t_0\in\mathbb{R}$, but it is necessary to supplement suitable boundary conditions on $\partial M$. Among the plethora of possible choices, we are mainly interested in those that entail the existence of unique advanced and retarded fundamental solutions associated to $P$ as well as of a two-point correlation function $\omega_2\in\mathcal{D}^\prime(M\times M)$, which allows to define a quantum state for the field $\Phi$ whose ultraviolet singular behaviour is of Hadamard form. 
	
	From a mathematical viewpoint the first problem has been addressed in \cite{Dappiaggi:2018jsq}, while the second one has been mainly discussed in \cite{Dappiaggi:2021wtr, Gannot:2018jkg}, although only on asymptotically anti-de Sitter spacetimes. It is important to stress that, in these works, the main task has always been the control of the ultraviolet behaviour of $\omega_2$, while no attention has been given to the possibility that infrared divergences might occur. This translates in the presence of an obstruction to the existence of $\omega_2$ as a well-defined distribution and, although this is abstractly known as a lurking possibility, to the best of our knowledge, it has not been sufficiently stressed that this might be directly connected to the choice of an underlying boundary condition. As mentioned in Section \ref{Sec: Introduction} this is the main goal of this paper and, to this end we start by recalling succinctly the procedure leading to the construction of ground states in the case in hand.

  	In order to make more crystal clear our message, we shall focus the attention only on boundary conditions of Robin type, although our conclusions can be drawn also for more general choices. Hence, starting from Equation \eqref{Eq: KG}, we consider the Fourier transform along the time-direction, namely
	$$\widehat{\Phi}(\omega,x)\doteq\int\limits_{\mathbb{R}}dt\, e^{i\omega t}\Phi(t,x),$$
	where $x$ stands for a choices of coordinates on $\Sigma$. Therefore Equation \eqref{Eq: KG} can be equivalently written as an eigenvalue problem
	\begin{equation}\label{Eq: Eigenvalue Problem}
		K\widehat{\Phi}=\omega^2\widehat{\Phi},
	\end{equation}
	where, denoting by $\Delta_h$ the Laplace-Beltrami operator built out of the Riemannian metric $h$, the operator $K$ reads
	\begin{equation}\label{Eq: Operator E}
		K=\beta\Delta_h-\frac{1}{2}h^{ij}(\partial_i\beta)\partial_j+\beta V.
	\end{equation}
	In studying Equation \eqref{Eq: KG}, the first step consists of establishing whether the Klein-Gordon operator $P$ admits unique advanced and retarded fundamental solutions $E_{adv}/E_{ret}\in\mathcal{D}^\prime(M\times M)$. As already mentioned, this is not the case unless one assigns suitable boundary conditions for the field $\Phi$ on $\partial M\simeq \mathbb{R}\times\partial\Sigma$. To this end, as discussed thoroughly in \cite{Dappiaggi:2018jsq}, it is convenient to read $K$ as a second order partial differential operator on $L^2(\Sigma)$, the space of square-integrable functions on a Cauchy surface $\Sigma$ with respect to the metric-induced induced volume form $\beta^{-1}d\mu_h$. In this way, it is possible to prove that admissible boundary conditions are in 1:1 correspondence with the self-adjoint extension of $K$ on $L^2(\Sigma)$. To make our case clear, among the plethora of possible choices, we restrict our attention only to those scenarios for which it is admissible a boundary condition that is {\em static} and of {\em Robin type}, namely there exists $\alpha\neq\alpha(t)\in C^\infty(\partial M)$ such that 
	\begin{equation}\label{Eq: Boundary Condition}
		\left.\Phi\right|_{\partial M}=\alpha\,\partial_n\Phi|_{\partial M},
	\end{equation}
	where $\partial_n$ denotes the derivative along the direction normal to the boundary. 
 
 Observe that in Equation \eqref{Eq: Boundary Condition}, the symbol $|_{\partial M}$ denotes the restriction both of $\Phi$ and of $\partial_n\Phi$ on the boundary. This operation is legitimate only if both $\Phi$ and $\partial_n\Phi$ are continuous on $\partial M$, a feature which is strongly dependent both on the choice of $V$ and of $g$. When the solution $\Phi$ to Equation \eqref{Eq: KG} is not regular enough, then all restrictions needs to be replaced by suitable trace maps, which we avoid defining in this paper in order to avoid unnecessary technical details. Hence we refer an interested reader to \cite{Dappiaggi:2018jsq} for the analysis of the general scenario.

 As a consequence of \cite[Prop. 19]{Dappiaggi:2018jsq}, for each boundary condition as in Equation \eqref{Eq: Boundary Condition} there exists a corresponding self-adjoint extension on $L^2(\Sigma)$ of the operator $K$ as in Equation \eqref{Eq: Operator E} which we denote by $K_\alpha$. In turn, this result combined with \cite[Thm. 30]{Dappiaggi:2018jsq}, entails that the operator $P$ admits unique advanced and retarded Green's operators, $E^\pm_\alpha$ whose associated integral kernels read $\mathcal{E}^-_\alpha(x,x^\prime)=\Theta(t-t^\prime)\mathcal{E}_\alpha(x,x^\prime)$ and $\mathcal{E}^+_\alpha(x,x^\prime)=-\Theta(t^\prime-t)\mathcal{E}_\alpha(x,x^\prime)$, where $\mathcal{E_\alpha}\in\mathcal{D}^\prime(M\times M)$ is such that, for all $f,f^\prime\in C^\infty_0(M)$,
	\begin{equation}\label{Eq: Causal Propagator}
		\mathcal{E}_\alpha(f,f^\prime)=\int\limits_{\mathbb{R}^2}dt dt^\prime\,\left(f(t),K^{-\frac{1}{2}}_\alpha\sin[K^{\frac{1}{2}}_\alpha(t-t^\prime)]f^\prime(t)\right)_{L^2(\Sigma)},	
	\end{equation}
	where $(,)_{L^2(\Sigma)}$ denotes the inner product of $L^2(\Sigma)$, while $K^{-\frac{1}{2}}_\alpha\sin[K^{\frac{1}{2}}_\alpha(t-t^\prime)]$ is defined in terms of the functional calculus for $K_\alpha$. Having established the existence of the advanced and retarded Green's operators associated to $P$, the next step in the construction of a quantum field theory consists of individuating $\omega_2\in\mathcal{D}^\prime(M\times M)$, the two-point correlation function of a quantum state. In the case in hand, and considering a boundary condition as per Equation \eqref{Eq: Boundary Condition}, this amounts to solving the system
	\begin{equation}\label{Eq: Defining Equation 2-pt}
		\left\{\begin{array}{l}
			\left(P\otimes\mathbb{I}\right)\omega_{2,\alpha}=0, \quad\left(\mathbb{I}\otimes P\right)\omega_{2,\alpha}=0,\\
			\omega_{2,\alpha}(f,f^\prime)-\omega_{2,\alpha}(f^\prime,f)=i\mathcal{E}_\alpha(f,f^\prime)\quad\forall f,f^\prime\in\mathcal{D}(M),
		\end{array}
		\right. 
	\end{equation} 
	where $\mathcal{E}_\alpha$ is as per Equation \eqref{Eq: Causal Propagator} and where the subscript $\alpha$ in $\omega_{2,\alpha}$ highlights the dependence of the two-point function from the boundary condition. Still exploiting the functional calculus for $K_\alpha$, Proposition 5.5. in \cite{Dappiaggi:2021wtr} entails that a solution to Equation \eqref{Eq: Defining Equation 2-pt}, corresponding to the two-point function of a ground state of Hadamard form exists if there exists in turn $\omega_{2,\alpha}\in\mathcal{D}^\prime(M\times M)$ which reads
	\begin{equation}\label{Eq: 2-pt function ground state}
		\omega_{2,\alpha}(f,f^\prime)=\int\limits_{\mathbb{R}^2}dt dt^\prime\,\left(f(t),K^{-\frac{1}{2}}_\alpha\exp[iK^{\frac{1}{2}}_\alpha(t-t^\prime)]f^\prime(t)\right)_{L^2(\Sigma)}.
	\end{equation}
While such expression appears to be innocuous, observe that the symmetric part of $\omega_{2,\alpha}$ reads formally
\begin{equation}\label{Eq: symmetric part 2-pt function ground state}
	\omega^S_{2,\alpha}(f,f^\prime)=\int\limits_{\mathbb{R}^2}dt dt^\prime\,\left(f(t),K^{-\frac{1}{2}}_\alpha\cos\left[K^{\frac{1}{2}}_\alpha(t-t^\prime)\right]f^\prime(t)\right)_{L^2(\Sigma)}.
\end{equation}
Yet, contrary to the antisymmetric part of $\omega_2$ which coincides up to multiplicative constants with Equation \eqref{Eq: Causal Propagator}, Equation \eqref{Eq: symmetric part 2-pt function ground state} is potentially ill-defined due to the $0$-modes of the operator $K_\alpha$ which entail that the integrand might fail to be integrable on account of the action of $K^{-\frac{1}{2}}_\alpha\cos\left[K^{\frac{1}{2}}_\alpha(t-t^\prime)\right]$. This potential singularity is of infrared type since it involves the low energy behaviour of the operator $P$ as in Equation \eqref{Eq: KG} and, in turn, it is directly dependent on the spectrum of $K_\alpha$, whose form depends explicitly on the underlying boundary condition as per Equation \eqref{Eq: Boundary Condition}. 

Having established at a generic and abstract level where is the source of a possible infrared singularity, in the next section, we discuss an explicit example that highlights it more concretely, making particularly evident the interplay with the underlying boundary condition. 
	

	\section{Infrared Divergences on Bertotti-Robinson Spacetime}\label{Sec: Bertotti-Robinson}
	
	In order to give an example of the feature outlined in the previous section, we consider as background $(M,g)$ a Bertotti-Robinson spacetime, which is a solution to the Einstein-Maxwell equation and it approximates the near horizon geometry of an extremal black hole with unit charge, see e.g. \cite{Bertotti, Robinson, Conroy2021aow, Ottewill:2012mq}. As a manifold, $M$ is globally diffeomorphic to $\textrm{CAdS}_2\times\mathbb{S}^2$, where $\textrm{CAdS}_2$ stands for the universal cover of the two-dimensional anti-de Sitter spacetime. In the following we start by considering a distinguished patch $(\widetilde{M},g)$ of $(M,g)$ which makes manifest the connection with a black hole spacetime, see \cite{Conroy2021aow}. Herein the line-element reads 	
	\begin{equation}\label{Eq: metric}
		ds^2 = -(\rho^2-1)dt^2 + (\rho^2-1)^{-1}d\rho^2 + d\Omega^2(\theta,\varphi),
	\end{equation}
	where $t\in\mathbb{R}$, $\rho\in (1,\infty)$ while $d\Omega^2(\theta,\varphi) = d\theta^2 + \sin^2\theta d\phi^2$ is the line element of the unit two-sphere. 
	On top of $(\widetilde{M},g)$ we consider a massive, real scalar field $\Psi:\widetilde{M}\to\mathbb{R}$, whose dynamics is ruled by 
	\begin{equation}\label{Eq: KG in BR}
		P_0\Psi=\left(\Box_g-m^2\right)\Psi=0,
	\end{equation}
	where $\Box_g$ is the D'Alembert wave operator built out of $g$ and where we are discarding any coupling to the underlying, scalar curvature since it is vanishing. Since $(\widetilde{M},g)$ is a static spacetime with a (conformal) timelike boundary at $\rho\to\infty$, {\it c.f.}  Equation \eqref{Eq: metric}, it is legitimate to look for a bi-distribution $\omega_2\in\mathcal{D}^\prime(\widetilde{M}\times\widetilde{M})$ which represents the two-point correlation function of a ground state. In the following we follow a procedure based on a mode decomposition which yields ultimately a counterpart both of Equation \eqref{Eq: Causal Propagator} and of Equation \eqref{Eq: 2-pt function ground state} in the case in hand. Let us therefore consider a solution $\Psi$ of Equation \eqref{Eq: KG in BR} which admits the following decomposition
	\begin{equation}
		\Psi (t,\rho,\theta,\phi) =\sum\limits_{l=0}^\infty\sum\limits_{p=-l}^l\int\limits_{\mathbb{R}} e^{-i\omega t}\psi_{\omega lp}(\rho) Y^p_l (\theta,\phi),
	\end{equation}
	where $Y^p_l(\theta,\varphi)$ is the standard spherical harmonic, eigenfunction of the Laplacian on the unit $2$-sphere, namely $\Delta_{\mathbb{S}^2}Y_{lp}(\theta,\varphi)=-l(l+1)Y_{lp}(\theta,\varphi)$. On account of Equation \eqref{Eq: KG in BR}, $\psi_{\omega lp}(\rho)$ satisfies the radial equation
	\begin{equation}\label{Eq: Radial Equation}
		(\mathbf{L}+\omega^2)\psi(\rho):= \bigg\{(\rho^2-1)\bigg[\frac{d}{d\rho}\bigg((\rho^2-1)\frac{d}{d\rho}\bigg)-l(l+1)-m^2\bigg]+\omega^2 \bigg\}\psi(\rho) = 0,
	\end{equation}
	which is a Sturm-Liouville problem with eigenvalue $-\omega^2$, see \cite{Zettl:2005}. Observe that we have suppressed the subscripts highlighting the dependence of $\mathbf{L}$ as well as of the solutions of the radial equation on the spectral parameters $\omega, l,p$ to avoid an unnecessarily heavy notation. A basis of solutions of Equation \eqref{Eq: Radial Equation} is given by the associated Legendre polynomials
	\begin{subequations}\label{Eq: basis of solutions}
	\begin{align}
		\psi_1(\rho)&=P^{i\omega}_\nu(\rho),\\
		\psi_2(\rho)&=Q^{i\omega}_\nu(\rho),\label{Eq: Secondary Solution}
	\end{align}
	\end{subequations}
	with $\nu=\frac{1}{2}(\sqrt{1+4l(l+1)+4m^2}-1)$, see \cite[\S 14.2]{NIST}. The next step consists of constructing a fundamental solution associated to the differential operator $\mathbf{L}$. Following the theory of Sturm-Liouville operators, see \cite{Zettl:2005}, first of all we allow $\omega$ to take any complex value and we look for solutions of Equation \eqref{Eq: KG in BR} that belong to $\mathcal{H}_1 := L^2((1,c),d\mu)$ or to $\mathcal{H}_\infty:= L^2((c,\infty),d\mu)$, with  $d\mu=(\rho^2-1)^{-1}d\rho$, while $c\in (1,\infty)$ is an arbitrary, but fixed number. 
	
	Since both $\psi_1$ and $\psi_2$ are smooth functions, it suffices to look for their asymptotic expansion in a neighborhood of $\rho=1$. Using \cite[\S 14.8]{NIST}, it turns out that $Q^{\pm i\omega}_\nu$, $P^{\pm i\omega}_\nu \in \mathcal{H}_1$ if $\pm\Imag(\omega) > 0$, while they both lie in $\mathcal{H}_\infty$ if and only if $0\leq \nu < \frac{1}{2}$, which is equivalent to $l=0$ and $m^2\in [-\frac{1}{4}, \frac{3}{4})$. 
	Whenever $\nu>\frac{1}{2}$, only $Q^{\pm i\omega}_\nu$ belongs to $\mathcal{H_\infty}$. 
	
	Hence, using the standard terminology proper of Sturm-Liouville problems summarized in \cite{Dappiaggi:2016fwc}, at $\rho=\infty$, we say that $\mathcal{Q}^\omega_\nu(\rho):=e^{\pi \omega}Q^{i \omega}_\nu(\rho)$ is the {\em principal solution}, since, as $\rho\to\infty$, it falls off to $0$ faster than any other solution of Equation \eqref{Eq: Radial Equation}. At the same time $\mathcal{P}^\omega_\nu(\rho):=\frac{\pi}{2i}\sinh^{-1}(\pi \omega) P^{i\omega}_\nu (\rho)$ is referred to as being a choice for \textit{one} secondary solution, see \cite[Def. 4.2]{Campos:2022byi} as well as \cite{Dappiaggi:2021wtr} for a discussion of the physical significance. We highlight that the $\omega$-dependent coefficients have also been chosen for later computational efficiency.
	
	Using Weyl classification of the behaviour of a Sturm-Liouville problem at the end points, we can infer that both $\rho=1$ and $\rho\to\infty$ are {\em limit circle} if $0<\nu<\frac{1}{2}$, while only $\rho\to\infty$ falls in this class if $\nu\geq\frac{1}{2}$. This entails that, in order to solve Equation \eqref{Eq: Radial Equation}, in addition to initial data we need to assign boundary conditions at the endpoints whenever they are limit circle. In the following, since we are more concerned about the r\^{o}le of the time-like boundary in the analysis of the underlying quantum theory, we consider only Dirichlet boundary conditions on the horizon ($\rho=1$), while, at $\rho\to\infty$, we assign Robin boundary conditions. Observe that, if we switch back to a fully-covariant description of the underlying field, Equation \eqref{Eq: metric} entails that the locus $\rho=1$ is an event horizon, hence a light-like hypersurface. This 
	
	This translates in the choice of the following solutions of Equation \eqref{Eq: Radial Equation} whenever $\Imag(\omega)\neq 0$:
	\begin{subequations}\label{Eq: Basis of Radial Functions}
	\begin{align}
		u_1(\rho)&:=\Theta(\Imag(\omega))P^{i\omega}_\nu(\rho)+\Theta(-\Imag(\omega))P^{-i\omega}_\nu(\rho),\\
		u_\infty(\rho)&:= \cos \gamma \, e^{\pi \omega}Q^{i \omega}_\nu(\rho) + \sin \gamma \, \frac{\pi}{2i}\sinh^{-1}(\pi \omega) P^{i\omega}_\nu (\rho), \qquad \gamma\in [0,\pi/2],
	\end{align}
	\end{subequations}
	where the Heaviside step function in $u_1$ helps us to take into account both $\Imag\omega$ positive and negative in a single expression, while the exponential 
	coefficients appearing in $u_\infty$ are chosen for future computational convenience. As a consequence of the choices above, it turns out that $u_\infty$ abides by Robin boundary conditions, namely
        \begin{equation}\label{Eq: Robin boundary conditions}
            \lim_{\rho\to\infty}(\cos\gamma \{ u_\infty(\rho),\mathcal{P}^\omega_\nu(\rho)\}+\sin\gamma \{u_\infty(\rho),\mathcal{Q}^\omega_\nu(\rho)\})=0,
        \end{equation}
    where $\{\cdot,\cdot\} := (\rho^2-1)W[\cdot,\cdot]$, $W$ being the Wronskian.

	\subsection{The radial Green function}\label{Sec: Radial Green Function}
	As a next step, denoting by $I$ the interval $(1,\infty)$, we look for $G_\gamma\in\mathcal{D}^\prime(I\times I)$, $\gamma\in [0,\frac{\pi}{2}]$, a distribution whose integral kernel $G_\gamma(\omega,\rho,\rho^\prime)\equiv G_\gamma(\rho,\rho^\prime)$ abides by
	\begin{equation}
		(\mathbf{L}\otimes \mathbb{I})G_\gamma(\rho,\rho') = (\mathbb{I}\otimes \mathbf{L})G_\gamma(\rho,\rho') = \delta(\rho-\rho'),
	\end{equation}
where $\mathbf{L}$ is the operator in Equation \eqref{Eq: Radial Equation} and where the subscript $\gamma$ highlights that we are looking for a fundamental solution which encodes Robin boundary conditions as per Equation \eqref{Eq: Robin boundary conditions}. Observe that, once more, we are suppressing for convenience of the notation any explicit reference on the dependence of $G_\gamma$ on $\omega$ as well as on $l,p$. 
	Using Equation \eqref{Eq: Basis of Radial Functions} and standard results of theory of second order ODEs, it holds that 	
	\begin{equation}\label{Eq: Radial Green Function}
		G_\gamma(\rho,\rho')=\frac{u_1(\rho_<)u_\infty(\rho^\prime_>)}{(\rho^2-1)W[u_\infty,u_1](\rho)},
	\end{equation}
	where 
	$$u_1(\rho_<)u_\infty(\rho^\prime_>)=\Theta(\rho-\rho')u_1(\rho')u_\infty(\rho)+\Theta(\rho'-\rho)u_1(\rho)u_\infty(\rho'),$$
	while $W[\cdot,\cdot]$ is the Wronskian between $u_1$ and $u_\infty$. As an example, for Dirichlet ($\gamma=0$) and Neumann boundary conditions ($\gamma=\frac{\pi}{2}$), Equation \eqref{Eq: Radial Green Function} boils down to
\begin{subequations}
	\begin{equation}
		G_0(\rho,\rho')=e^{\pi\omega}P^{-i\omega}_\nu(\rho_<)Q^{i\omega}_\nu(\rho_>),
	\end{equation}	
	\begin{equation}
		G_\frac{\pi}{2}(\rho,\rho')=\frac{-i\pi}{2\sinh\pi\omega}P^{-i\omega}_\nu(\rho_<)P^{i\omega}_\nu(\rho_>),
	\end{equation}
\end{subequations}	
	
\noindent while, for $\gamma\in(0,\pi/2)$, \textit{i.e.} for Robin boundary conditions, it holds that	
	\begin{equation}\label{Eq: Radial Green Function 2}
		G_\gamma(\rho,\rho')=\frac{\cos\gamma G_0(\rho,\rho') + \sin \gamma G_\frac{\pi}{2}(\rho,\rho')}{\cos\gamma +\sin\gamma}.
	\end{equation} 

Observe that if we reinstate the $\omega$-dependence in the radial Green function, it holds that, for all $\gamma\in[0,\frac{\pi}{2}]$, $G_\gamma(-\omega,\rho,\rho') = \overline{G_\gamma(\omega,\rho,\rho')}$ whenever $\omega\in\mathbb{R}$.  We remark that given $G_\gamma$ there is always the freedom to add a bi-solution of the radial equation $\mathbf{L}G_0=0$ which abides by the Robin boundary condition parameterized by $\gamma\in [0,\frac{\pi}{2}]$. In our construction this freedom amounts to the possibility of making different choices of the secondary solution in Equation \eqref{Eq: Secondary Solution}. The consequences of this leeway have been thoroughly investigated in \cite{Campos:2022byi,Campos:2022waj} and we shall not delve further into them.
	
	\subsection{Resolution of the Identity}
	Having established an expression for the radial Green function, in order to construct the two-point correlation function for the ground state of the Klein-Gordon field as in Equation \eqref{Eq: KG in BR} endowed with Robin boundary conditions, we need to individuate a resolution of the identity written in terms of the Green function $G_\gamma$ as in Equation \eqref{Eq: Radial Green Function 2}. Following \cite[Ch. 7]{greenBook}, reinstating the explicit dependence on $\omega$ and recalling that we are allowing $\omega\in\mathbb{C}$, it holds that
	
	\begin{equation}\label{Eq: Contour Integral}
		-(\rho^2-1)\delta(\rho-\rho')=\frac{1}{2\pi i}\int_{C_\infty}G_\gamma(\lambda, \rho,\rho')d\lambda, \quad \lambda=\omega^2,
	\end{equation}
	where $C_\infty$ is a suitable ``Pac-Man" contour in the complex $\lambda$-plane. Using Jordan's lemma, the right hand side of Equation \eqref{Eq: Contour Integral} can be shown to abide by the following chain of identities:
	\begin{align}\label{Eq: Resolution Right Hand Side}
		\frac{1}{2\pi i}\int_{C_\infty}G_\gamma(\lambda, \rho,\rho')d\lambda  \overset{\lambda=\omega^2}{=} \frac{1}{\pi i}\int_\mathbb{R} d\omega\, \omega\, G_\gamma(\omega,\rho,\rho') &=  \frac{1}{\pi i}\int_0^\infty d\omega \, \omega \,[G_\gamma(\omega,\rho,\rho')-G_\gamma(-\omega,\rho,\rho')]=\\
        &=\frac{2}{\pi }\int_0^\infty d\omega \, \omega \, \Imag[G_\gamma(\omega,\rho,\rho')].
	\end{align}
	Introducing the function $R_\gamma(\omega,\rho,\rho^\prime):=\frac{2}{\pi}\Imag[G_\gamma(\omega,\rho,\rho^\prime)]$, by comparison with Equation \eqref{Eq: Contour Integral}, it holds that
        \begin{equation}\label{Eq: Resolution of the Identity}
            \int_\mathbb{R} d\omega \, \omega \, R_\gamma(\omega,\rho,\rho') := (\rho^2 -1)\delta(\rho-\rho'),
        \end{equation}
        which is a real-valued odd function in $\omega$, {\it cf.} Equation \eqref{Eq: Resolution Right Hand Side}. Using Equation \eqref{Eq: Radial Green Function 2}, one can infer that
	\begin{subequations}
	\begin{align}\label{Eq: R-functions}
		&R_0(\omega,\rho,\rho')=\frac{i}{\pi}\big(e^{\pi\omega}P^{-i\omega}_\nu(\rho)Q^{i\omega}_\nu(\rho')-e^{-\pi\omega}P^{i\omega}_\nu(\rho)Q^{-i\omega}_\nu(\rho')\big),\\
		&R_{\frac{\pi}{2}}(\omega,\rho,\rho')=\frac{1}{2\sinh\pi\omega}(P^{-i\omega}_\nu(\rho)P^{i\omega}_\nu(\rho')+P^{i\omega}_\nu(\rho)P^{-i\omega}_\nu(\rho')),\\
		&R_\gamma(\omega,\rho,\rho')=\frac{\cos\gamma R_0(\omega,\rho,\rho') + \sin \gamma R_\frac{\pi}{2}(\omega,\rho,\rho')}{\cos\gamma +\sin\gamma}, \qquad \gamma\in\big( 0,\,\frac{\pi}{2}\big).
	\end{align}
\end{subequations}
	
\subsection{Two-point correlation function of the ground state}\label{Sec: 2-pt correlation function of the ground state}
\begin{figure}[h!]
        \centering
        \includegraphics[width=0.7\textwidth]{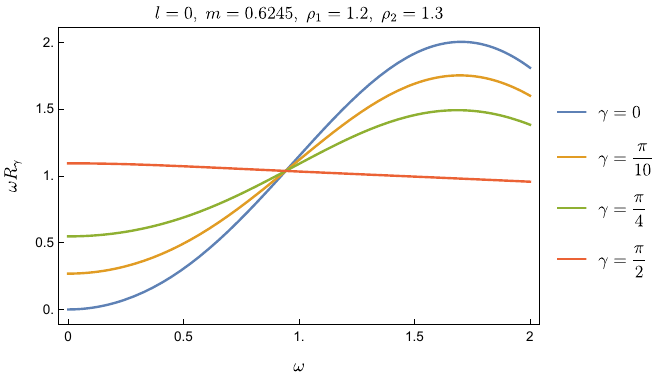}
        \caption{The radial function $R_\gamma$ as in Equation \eqref{Eq: R-functions} where we set $l=0$. To highlight the behaviour near $\omega=0$, we have considered the product $\omega R_\gamma$ as a function of $\omega$, the other parameters being fixed. The plot shows that only for $\gamma=0$ $\omega R_\gamma$ tends to $0$ as $\omega\to 0$, all other cases displaying therefore a singular behaviour.}
        \label{plot1}
    \end{figure}

Having individuated the necessary building blocks, we are in a position to address the question of the construction of a two-point correlation function for the ground state of a Klein-Gordon field as per Equation \eqref{Eq: KG in BR} with Robin boundary conditions. In other words, denoting by $(M,g)$ the underlying Bertotti-Robinson spacetime as per Equation \eqref{Eq: metric}, we are looking in the first place for $\omega_{2,\gamma}\in\mathcal{D}^\prime(M\times M)$, the subscript $\gamma$ highlighting the dependence on the Robin boundary condition, such that 
$$(P\otimes \mathbb{I})\omega_{2,\gamma} = (\mathbb{I}\otimes P)\omega_{2,\gamma} = 0,$$ 
while its antisymmetric part is such that, working at the level of integral kernels
	\begin{equation}\label{Eq: antisymmetric part 2-pt function}
	\omega_{2,\gamma}(x,x')-\omega_{2,\gamma}(x',x) = i\mathcal{E}_\gamma(x,x'),
\end{equation}
where $\mathcal{E}_\gamma\in\mathcal{D}'(\mathcal{M}\times\mathcal{M})$ is the causal propagator, namely the difference between the retarded and the advanced fundamental solutions of the operator $P$ as in Equation \eqref{Eq: KG in BR} supplemented with Robin boundary conditions -- see also Section \ref{Sec: General Construction}. In turn $\mathcal{E}_\gamma$ abides by the following initial value problem:
	\begin{flalign}
		(P\otimes \mathbb{I})\mathcal{E}_\gamma = (\mathbb{I}\otimes P)\mathcal{E}_\gamma = 0,\\
		\mathcal{E}_\gamma|_{\Sigma_t \times \Sigma_t} = 0, \qquad \partial_t \mathcal{E}_\gamma|_{\Sigma_t \times \Sigma_t} = \delta_{\Sigma_t},
	\end{flalign}
	where $\Sigma_t$ is a generic Cauchy surface at constant time $t\in\mathbb{R}$, while $\delta_{\Sigma_t}$ is the Dirac delta on $\Sigma_t$. Following \cite[Ch. 2.3]{DeSouzaCampos2022wsp}, a solution to this initial value problem can be written as 
	\begin{equation}\label{Eq: Causal Propagator_2}
		\mathcal{E}_\gamma(x,x') =\lim_{\varepsilon\to 0^+} \sum_{l=0}^\infty\sum_{p=-l}^{+l}\int_\mathbb{R} d\omega \sin{(\omega(t-t'-i\varepsilon))} Y_l^p(\theta,\phi) \overline{Y_l^p(\theta',\phi')} R_\gamma(\omega,\rho,\rho'),
	\end{equation}
	where $R_\gamma(\omega,\rho,\rho')$ ought to satisfy the integral identity
	\begin{equation}
		\int_\mathbb{R} d\omega \, \omega \, R_\gamma(\omega,\rho,\rho') = (\rho^2 -1)\delta(\rho-\rho').
	\end{equation}
	This is nothing but Equation \eqref{Eq: Resolution of the Identity} which justifies why we have used the symbol $R_\gamma$. Keeping in mind Equation \eqref{Eq: antisymmetric part 2-pt function} as well as Equation \eqref{Eq: Causal Propagator_2}, we have all ingredients to write the formal expression of the two-point correlation function of the ground state associated to a Klein-Gordon field with Robin boundary conditions on a Bertotti-Robinson spacetime, namely, using \cite[Th. 2.22]{DeSouzaCampos2022wsp}:
    \begin{equation}
        w_{2,\gamma}(x,x') =\lim_{\varepsilon\to 0^+} \sum_{l=0}^\infty\sum_{p=-l}^{+l}\int_\mathbb{R} d\omega\, \Theta(\omega) e^{-i\omega(t-t'-i\varepsilon)} Y_l^p(\theta,\phi) \overline{Y_l^p(\theta',\phi')}R_\gamma(\omega,\rho,\rho'),
    \end{equation}
    which can be rewritten in a more compact form, summing over $m$ and using \cite[\S 14.30.9]{NIST}, as
	
	\begin{equation}\label{Eq: 2-pt function Patch Hor}
		w_{2,\gamma}(x,x') = \lim_{\varepsilon\to 0^+}\sum_{l=0}^{\infty}\int_0^\infty d\omega e^{-i\omega(t-t'-i\varepsilon)}\frac{2l+1}{4\pi} P_l (\cos\Gamma(\theta,\theta',\phi,\phi')) R_\gamma(\omega,\rho,\rho'),
	\end{equation}
	where $P_l$ is the Ferrers function of the first kind while $\Gamma:\mathcal{S}^2 \times \mathcal{S}^2 \to \mathbb{R}$ is the geodesic distance on the unit 2-sphere. We highlight that we have called $w_{2,\gamma}$ a formal expression since if we expand $R_\gamma$ near $\omega=0$, for both $\gamma=0$ and $\gamma \in (0,\frac{\pi}{2}]$, the asymptotic behaviour of $R_\gamma$ reads -- see Figure \ref{plot1}
    \begin{equation}
        |R_0(\omega,\rho,\rho')| \overset{\omega \to 0}{\sim} \omega,\qquad  |R_\gamma(\omega,\rho,\rho')| \overset{\omega \to 0}{\sim} \omega^{-1}.
    \end{equation}
    Hence the expression in Equation \eqref{Eq: 2-pt function Patch Hor} identifies a well defined distribution only for $\gamma = 0$, namely Dirichlet boundary conditions, while for $\gamma\in(0,\frac{\pi}{2}]$ an infrared divergence occurs, therefore Equation \eqref{Eq: 2-pt function Patch Hor} cannot identify the two-point correlation function of a ground state.

    \subsection{A different scenario: the Poincar\'e patch}
    
\begin{figure}[h!]
        \centering
        \includegraphics[width=0.7\textwidth]{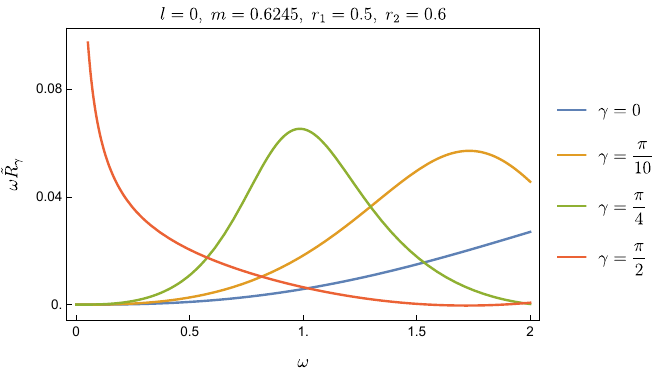}
        \caption{The radial function $\Tilde{R}_\gamma$ as in Equation \eqref{Eq: TildeR} where we set $l=0$. To highlight the behaviour near $\omega=0$, we have considered the product $\omega \Tilde{R}_\gamma$ as a function of $\omega$, the other parameters being fixed. The plot shows that the function tends to $0$ as $\omega\to 0$ for all boundary conditions except for $\gamma=\frac{\pi}{2}$.}
        \label{plot2}
    \end{figure}

	In the previous section we have shown that infrared divergences can occur in a Bertotti-Robinson spacetime when the underlying Klein-Gordon field is not endowed with Dirichlet boundary conditions. In the following, we highlight that the correspondence between the existence of a singular behaviour at large distances and the choice of a boundary condition is highly dependent on the underlying geometry. To achieve this goal, we consider a different patch of the Bertotti-Robinson spacetime, denoted by $(\widetilde{M}_1,g_1)$. It has been already analyzed in \cite{Campos:2022waj} and it can be realized starting from Equation \eqref{Eq: metric} and switching from the coordinates $(t,\rho)$ to $(\tau, r)$ by means of the transformation \cite{Ottewill:2012mq}
    \begin{equation}
        \tau = \frac{(\rho^2-1)^{1/2}\sinh t}{\rho + (\rho^2-1)^{1/2}\cosh t}, \qquad r=\frac{1}{\rho+(\rho^2-1)^{1/2}\cosh t}.
    \end{equation}
    The line element in Equation \eqref{Eq: metric} becomes
    \begin{equation}\label{Eq: metric_new_patch}
        ds^2 = \frac{1}{r^2}(-d\tau^2+dr^2+r^2 d\Omega^2),
    \end{equation}
	where the new coordinates can be assigned the following domain: $\tau\in\mathbb{R}$ while $r\in (0,\infty)$. Subsequently, on top of $\widetilde{M}_1$, we consider a massive, real, Klein-Gordon field $\widetilde{\Psi}:\widetilde{M}_1\to\mathbb{R}$ whose dynamics is ruled by Equation \eqref{Eq: KG in BR} where the D'Alembert wave operator is built out of the metric in Equation \eqref{Eq: metric_new_patch}. In particular we are interested in constructing the two-point correlation function of the ground state associated to $\widetilde{\Psi}$ and we can follow the same step-by-step construction outlined in the previous section. Here we will not dwell into the details which have been already accounted for in \cite{Campos:2022waj} and we report the main ingredients and results. We observe that in the last cited paper the main interest lies in the analysis of the response function of a Unruh-de Witt detector rather than in the identification of infrared singularities. As in the previous section we consider a mode decomposition 

    \begin{equation}
        \widetilde{\Psi} (\tau,r,\theta,\phi) =\sum\limits_{l=0}^\infty\sum\limits_{p=-l}^l\int\limits_{\mathbb{R}} e^{-i\omega \tau}\widetilde{\psi}_{\omega lp}(r) Y^p_l (\theta,\phi),
    \end{equation}
    where $\widetilde{\psi}_{\omega lp}$ satisfies the radial equation
    \begin{equation}
        (\mathbf{\Tilde{L}}+\omega^2)\widetilde{\psi}_{\omega lp}(r):=\bigg[  \frac{d^2}{dr^2} - \frac{l(l+1)+m^2}{r^2} + \omega^2  \bigg]\widetilde{\psi}_{\omega lp}(r).
    \end{equation}
    A basis of solutions of this equation can be written in terms of Bessel functions of the first kind and, using once more the language of Sturm-Liouville problems, the primary and secondary solutions read
    \begin{equation}\label{Eq: Basis of solutions patch 2}
        \Tilde{\mathcal{P}}(r) =\sqrt{r}J_\eta (\omega r) , \qquad  \Tilde{\mathcal{S}}(r)=-\omega^{2\eta}\sqrt{r}J_{-\eta}(\omega r),
    \end{equation}
    where $\eta =\frac{1}{2}\sqrt{1+4l(l+1)+4m^2}$ and where, for simplicity of the notation, we have dropped the subscript highlighting the dependence on $\omega, l,p$. As discussed in \cite[Sec. 2]{Campos:2022waj}, Robin boundary conditions, parameterized by $\gamma \in [0,\frac{\pi}{2}]$, can be imposed whenever $l=0$ and $m^2\in[-\frac{1}{4},\frac{3}{4})$ exactly as in the previous section. Using Equation \eqref{Eq: Basis of solutions patch 2}, it turns out that the two-point correlation function for the ground state in this specific patch reads formally
    \begin{equation}\label{Eq: 2-pt function Patch 1}
		\widetilde{w}_{2,\gamma}(x,x') = \lim\limits_{\varepsilon\to 0^+}\sum_{l=0}^{\infty}\int_0^\infty d\omega e^{-i\omega(\tau-\tau'-i\varepsilon)}\frac{2l+1}{4\pi} P_l (\cos\Gamma(\theta,\theta',\phi,\phi')) \Tilde{R}_\gamma(\omega,r,r'),
	\end{equation}
    where, similarly to Equation \eqref{Eq: 2-pt function ground state} we have taken the sum over all admissible values of $p$ and where 
    \begin{equation}\label{Eq: TildeR}
        \Tilde{R}_\gamma(\omega,r,r') = \frac{\sqrt{rr'}(\cos(\gamma)J_\eta(\omega r)+\sin(\gamma)\omega^{2\eta}J_{-\eta}(\omega r))(\cos(\gamma) J_\eta(\omega r')+\sin(\gamma)\omega^{2\eta}J_{-\eta}(\omega r'))}{2(\sin^2(\gamma)\omega^{4\eta}+\sin(2\gamma)\cos(\pi\eta)\omega^{2\eta}+\cos^2(\gamma))}.
    \end{equation}
    If we expand $\Tilde{R}_\gamma$ near $\omega=0$, for both $\gamma \in [0,\frac{\pi}{2})$ and $\gamma=\frac{\pi}{2}$, the asymptotic behaviour of $\Tilde{R}_\gamma$ reads
    \begin{equation}
        |\Tilde{R}_\gamma(\omega,r,r')| \overset{\omega \to 0}{\sim} \omega^{2\eta},\qquad  |\Tilde{R}_{\frac{\pi}{2}}(\omega,r,r')| \overset{\omega \to 0}{\sim} \omega^{-2\eta}.
    \end{equation}
    Since we are considering mass values in the range $m^2 \in [0,\frac{3}{4})$, then the parameter $\eta\in [\frac{1}{2},1)$ which entails that the right hand side of Equation \eqref{Eq: 2-pt function Patch 1} identifies a well-defined distribution whenever $\gamma\in [0,\frac{\pi}{2})$ while for $\gamma=\frac{\pi}{2}$, namely only for Neumann boundary conditions, an infrared divergence occurs in the integrand, see Figure \ref{plot2}. Therefore one cannot conclude that Equation \eqref{Eq: 2-pt function Patch 1} identifies the two-point correlation function of a ground state when $\gamma=\frac{\pi}{2}$.

	\section{Conclusions}
	
	In this short paper, we have highlighted that, in the construction of the two-point correlation functions for a free, Klein-Gordon field on globally hyperbolic manifolds with a timelike boundary, one might face the occurrence of infrared singularities in the construction of the two-point correlation function of the ground state even when considering simple boundary conditions such as those of Robin type. Although this pathology is a well-known feature especially in two-dimensional globally hyperbolic spacetimes with an empty boundary, we have shown that it can occur even in higher dimensions, {\it e.g.} in a Bertotti-Robinson spacetime, and that it is strictly linked to the choice of boundary conditions for the underlying field theory. It is interesting to remark that, since this phenomenon occurs for ground states, then it is also present for thermal/KMS states whose infrared behaviour is more singular due to the contribution of the Bose factor in the mode decomposition. Hence, one might envisage scenarios, behaving similarly to free Bosonic quantum field theories in three-dimensional globally hyperbolic spacetimes with an empty boundary, for which the ground state exists while the thermal counterpart displays infrared singularities. Finally, our work sets the ground for multiple future analyses among which we reckon that the following are especially worth mentioning:
	\begin{itemize}
		\item investigating the occurrence of infrared singularities when we consider a more general class of boundary conditions such as those of Wentzell type \cite{Dappiaggi:2018pju, Dappiaggi:2022dwo} and those associated to a dynamic wall \cite{Obadia:2001hj, WanMokhtar:2022npv},
		\item a comparison of the phenomenon highlighted in this work with the freedom of choosing the secondary solution in a Sturm-Liouville problem as highlighted in \cite{Campos:2022byi}.
	\end{itemize}

	\section{Acknowledgements}
	The work of L.C. is supported by a postdoctoral fellowship of the Department of Physics of the University of Pavia, while that of L.S. by a PhD fellowship of the University of Pavia.


\end{document}